\documentstyle[12pt,aasms4]{article}

\begin{document}

\title{ISOCAM Mid-Infrared Imaging\\of the Quiescent Spiral
Galaxy 
NGC 7331\footnote{Based on observations made with ISO, an ESA
project with instruments funded by ESA Member States and
with the participation of ISAS and NASA.}}
\author{Beverly J. 
Smith\footnote{Now at CASA, University of Colorado,
Box 389, Boulder CO 80309}}
\affil{IPAC/Caltech, MS 100-22, Pasadena CA  91125}

\vskip 0.2in
To appear in the Astrophysical Journal

\begin{abstract}

Using the mid-infrared camera (ISOCAM) on the Infrared Space
Observatory (ISO),
the Sb LINER galaxy NGC 7331 has
been imaged in two broadband and four narrowband filters
between 6.75 and 15 $\mu$m.   These maps show a prominent
circumnuclear ring of radius 0\farcm25 $\times$ 0\farcm75 
(1.1 $\times$ 3.3 kpc)
encircling an extended central source.
The 7.7 and 11.3 $\mu$m dust emission features are strong
in this galaxy, contributing $\sim$1/3 of the total IRAS
12 $\mu$m broadband flux from this galaxy.
In contrast to starburst galaxies,
the 15 $\mu$m continuum is weak in NGC 7331.
The mid-infrared spectrum 
does not vary dramatically with position in this
quiescent galaxy, 
showing neither large-scale destruction of
the carriers of the emission bands or
a large increase in the 15 $\mu$m continuum
in the star forming ring.
In the 
bulge, there is some enhancement of 
the 6.75 $\mu$m flux,
probably because of 
contributions from 
photospheric light,
however, 
the 
11.3 $\mu$m dust feature is also seen,
showing additional emission from
interstellar or circumstellar dust. 

\end{abstract}


\section{Introduction}

The standard picture of interstellar dust has
changed dramatically in recent years, with 
the realization that molecule-sized dust grains
transiently heated to very high temperatures may contribute significantly
to the infrared emission from the interstellar medium (\cite{s84};
\cite{bp88}).  
The characteristic set of mid-infrared emission features 
at 3.3, 6.2, 7.7, 8.6,
and 11.3 $\mu$m, commonly known as the
Unidentified Infrared Bands (UIBs), are likely caused by very small grains. 
Polycyclic aromatic hydrocarbons
(PAHs) are most often cited 
as the origin of these features
(\cite{lp84});
other possible carriers
are hydrogenated amorphous carbons (\cite{b87}),
quenched carbonaceous compounds (\cite{sak84}), and coal grains
(\cite{pap89}).
Before the advent of the Infrared Space Observatory (ISO;
\cite{k96}), 
ground-based 
spectroscopy 
revealed the presence of UIBs in luminous starburst galaxies
(e.g., \cite{r91}),
however,
the strength of these features in lower luminosity
galaxies was unknown.
IRAS broadband studies showed an increase in the 12 $\mu$m/60 $\mu$m
ratio with decreasing far-infrared luminosity L$_{FIR}$
(\cite{s87}; \cite{hrs91}).  This suggests
that small grains dominate the emission in the IRAS 12 $\mu$m bandpass
and are systematically destroyed in 
environments
with
strong UV radiation 
fields, and so the UIB features should be strong in lower luminosity 
galaxies.
This conclusion, however, could not be verified spectroscopically.

The higher sensitivity of ISO has now made it
possible to study very small grains in lower luminosity galaxies
as well as
a variety of environments in starburst galaxies.  
With the ISO Camera 
(ISOCAM; \cite{c96}), spatial as well as spectral information
is obtained, providing important clues to the distribution
of small grains within galaxies.
For example,
imaging
spectroscopy 
of the starburst Antennae galaxies
with the Circular Variable Filter (CVF) mode
of CAM
showed differences in the mid-infrared spectra
of starburst vs more quiescent regions (\cite{v96}); areas with
large populations of OB stars have 
more emission in the 15 $\mu$m continuum relative to
the UIBs.  
For fainter galaxies, where CVF imaging
is not feasible, the strength and distribution of the UIBs
can be estimated indirectly,
using broadband ISO mid-infrared colors
(e.g., \cite{sau96}; \cite{h96}).

The narrowband ISOCAM filters provide a better method of
tracing very small grains 
in fainter galaxies, because they isolate the 
UIB lines and give a better measure of the continuum
than broadband filters.
In this paper, both broad- and narrow-band ISOCAM
mid-infrared 
images of the Sb galaxy NGC 7331 are presented.
NGC 7331 is a 
nearby
(15 Mpc for H$_o$ = 75 km s$^{-1}$ Mpc$^{-1}$)
relatively undisturbed spiral
galaxy (see Figure 1 (Plate L1)).
It has
L$_{FIR}$
$\sim$ 2 $\times$ 10$^{10}$ L$_{\sun}$ (\cite{r88}), similar
to that of the Milky Way 
(\cite{c86}), 
a weak LINER-like nucleus (\cite{k83a}), and a
post-starburst population
in the bulge (\cite{ot96}).
Radio continuum (\cite{c94}) and H$\alpha$+[N~II] 
imaging 
(\cite{k83b}; \cite{p89})
show a faint 
nucleus
surrounded by weak diffuse bulge emission, 
implying
a low star formation rate in the bulge.
The bulge is encircled by 
prominent spiral arms in a ring-like structure.
The bulge appears to be depleted of both HI
(\cite{b81}) and molecular gas 
(\cite{von96}; \cite{ts97}), compared to the surrounding
spiral arms.  
In the mid-and far-infrared, IRAS could not
resolve the bulge from the arms;
this is also the case for Kuiper Airborne Observatory
far-infrared maps 
(\cite{sh96}).
The new ISO images provide not just higher spatial
resolution and sensitivity than IRAS, but also 
spectral information.  By imaging in narrowband filters,
we are able to isolate emission in the dust features from continuum
emission,
and directly detect the UIB features
in a non-starburst galaxy and determine their spatial
distribution.

\section{Observations and Data Reduction}

The ISO observations were made on
1996 December 25, 26, and 30 using the 32$\times$32 element ISOCAM array
on the 0.6m ISO telescope.
Six different filters were used:
the broadband LW2 
($\lambda$$_0$ = 6.75 $\mu$m; $\Delta$$\lambda$ = 3.5 $\mu$m) and
LW3 
($\lambda$$_0$ = 15 $\mu$m; $\Delta$$\lambda$ = 6 $\mu$m) filters
and the narrowband LW6 
($\lambda$$_0$ = 7.75 $\mu$m; $\Delta$$\lambda$ = 1.5 $\mu$m), LW7
($\lambda$$_0$ = 9.62 $\mu$m; $\Delta$$\lambda$ = 2.2 $\mu$m), LW8
($\lambda$$_0$ = 11.4 $\mu$m; $\Delta$$\lambda$ = 1.3 $\mu$m), and LW9
($\lambda$$_0$ = 15 $\mu$m; $\Delta$$\lambda$ = 2 $\mu$m) filters.
The observed region is marked in Figure 1 (Plate L1).
A 7 $\times$ 2 mosaic was made, using 
3$''$ pixels 
and an overlap
of 45$''$, giving a total field of view of                     
135$''$ $\times$ 350$''$.  The major axis of the raster is aligned at
a position angle of 80$^{\circ}$, perpendicular  
to the major axis of the galaxy, to give
sufficient off-galaxy sky to determine             
the background.  Along the major axis of the galaxy,
the coverage is sufficient to map the inner spiral arms
of NGC 7331.
Twelve five second exposures were made per raster position.
For stabilization, before these observations 
a series of additional exposures were made, with counts ranging from 19 to 67.

After standard ISOCAM pipeline reduction,
additional data reduction was accomplished using the CAM Interactive
Analysis (CIA) software\footnote{CIA is a 
joint development by the ESA Astrophysics Division and the ISOCAM 
Consortium led by the ISOCAM PI, C. Cesarsky, Direction des Sciences de  
la Mati\`ere, C.E.A., France.} as well as software developed at IPAC.
Deglitching was done automatically
with the Multi-resolution Median Transform routine
(\cite{s95}) and by hand with an interactive deglitching routine
developed at IPAC, and vignetted pixels on the edges of the array
were masked.
Transient effects (\cite{c96}) were corrected for 
using the latest version of the IPAC simple analytic 
transient removal software (\cite{s97}), which models both
upwards and downwards transients.
Flat fields were created using the median of the off-galaxy raster
positions, and the library darks were used.  
After dark subtraction and flat fielding, the individual frames
were 
mosaicked and the sky subtracted, and
the standard
ISO library calibration factors were used.
The calibration uncertainty is 
estimated to be $\le$30$\%$.

\section{The Mid-Infrared Morphology}

The six ISOCAM images 
are presented in Figures 2a to 2f (Plates L2a-f).
A prominent ring is seen at all wavelengths, with a diameter
of 0\farcm5 $\times$ 1\farcm5 (2.2 kpc $\times$ 6.5 kpc).
Inside this
ring, a central source is visible. This source is resolved
with respect to the ISO beam,
with an observed FWHM size of $\sim$7$''$ $-$ 9$''$ along the
major axis.  
Outside of the ring, secondary arcs are visible.
For comparison, the red continuum 6435$\AA$ narrowband image
and the H$\alpha$+[N~II] map from Pogge (1989) 
are shown in Figures 3a and 3b (Plates L3a and L3b),
along with the LW3 image (Figure 3c) (Plate L3c).
In Figures 3d, 3e, and 3f (Plates L3d-f), 
the H$\alpha$+[N~II] map,
the radio
continuum map of \markcite{c94}Cowan et al. (1994), and the CO 
map of \markcite{ts97}Tosaki $\&$ Shioya (1997), respectively,
are overlaid on the LW3 map.
These maps show
a good correspondence between
the mid-infrared structure and features seen at other wavelengths.
There is a good match between the mid-infrared and
H$\alpha$+[N~II] peaks (Figure 3d), however, the ring is 
much more prominent in the mid-infrared than in H$\alpha$+[N~II].
This is particularly true for the western side of the ring which
is quite faint in the optical, probably because of
high extinction in this region.
In the radio continuum and CO maps as well as in the mid-infrared, these
obscured H~II regions are bright.
The optically-prominent H~II regions in the spiral arms outside
of the ring
are also present in the mid-infrared maps,
but at much lower flux levels (see Figures 2, 3c, and 3d).

Although the NGC 7331 mid-infrared morphology 
resembles
that seen in the Sb galaxy M31 by IRAS
(\cite{h84}), 
there are some important differences between these galaxies.
M31 is an even more quiescent galaxy than NGC 7331,
with L$_{FIR}$
$\sim$25 times lower than NGC 7331 and
a much colder F$_{60}$/F$_{100}$ color temperature
(\cite{r88}).
Furthermore, the M31 ring is $\sim$3 times larger than
the NGC 7331 ring.
NGC 7331 may also resemble the Milky Way in mid-infrared morphology.
The Galaxy
is believed to contain a molecular ring of radius $\sim$5$-$6 kpc 
(\cite{sanders84},
\cite{bron88}),
which may also have an infrared counterpart (\cite{sodroski97}).
IRAS maps show a central mid-infrared peak in the
Milky Way with a similar size scale as the source in NGC 7331 (\cite{g84}).

\section{The Mid-Infrared Spectral Energy Distribution }

The mid-infrared spectrum of NGC 7331 is given
in 
Figure 4a.
The LW6 (7.75 $\mu$m) and LW8 (11.4 $\mu$m)
fluxes are enhanced relative to the fluxes in
the narrowband continuum filters LW7 (9.62 $\mu$m)
and LW9 (15 $\mu$m), indicating that the 
7.7 $\mu$m
and 11.3 $\mu$m UIB features
are present in NGC 7331.
The spectrum of the nearby sky is featureless,
consistent with the \markcite{r96}Reach
et al. (1996) zodiacal
spectrum (Figure 4b).
The lack of UIB emission in the off-galaxy positions 
confirms that the excesses seen in the LW6 and LW8 filters 
at the galaxy are real and extragalactic, rather than
being due to calibration errors or foreground dust.
In Figure 4b, 262K and 285K blackbody curves 
are also shown.
The first temperature was found by \markcite{r96}Reach
et al. (1996) for their data, 
while the latter is the best fit to the COBE DIRBE
data for a 3$^{\circ}$ region centered on NGC 7331,
interpolated to the ISO observation date (\cite{r97}).
The ISO sky fluxes agree with predictions based on the DIRBE data
to within 15$\%$.

The total IRAS broadband 12 $\mu$m flux density for NGC 7331
is 4.2 Jy (\cite{r88}), compared to $\sim$2 Jy
seen by ISO, so $\sim$1/2 of the total mid-infrared flux
of NGC 7331 is contained within the mapped region.
Integrating over the 8 $-$ 15 $\mu$m IRAS 12 $\mu$m band, 
the UIBs contribute $\sim$1/3 of the 
total IRAS 12 $\mu$m flux from NGC 7331.

In Figure 5, the NGC 7331 spectrum is compared to 
the ISOCAM CVF data for two
regions in the Antennae galaxies (from \cite{v96}): a region containing
a powerful starburst (Knot A), and a more quiescent region (the
nucleus of NGC 4038).  
The NGC 7331 spectrum
agrees
with that for the NGC 4038 nucleus but 
not the starburst region Knot A.  
NGC 7331 does not have the
steep continuum spectrum seen in starbursts,
but rather has a flatter continuum.
The difference between the LW3 and LW9 fluxes
is probably due to 
contributions to the broader LW3 filter
from 
the 12.8 $\mu$m [Ne~II] line, the 12.7 $\mu$m UIB,
and emission beyond 16.5 $\mu$m.

Individual spectra for the NGC 7331 ring and a
16\farcs5 $\times$ 12\farcs0 diameter region (1.2 kpc $\times$ 0.9 kpc)
around the 
bulge are
given in Figures 6a and 6b, 
using 
an LW3 surface brightness cutoff of 
12 MJy sr$^{-1}$.  
In Figure 6a the spectrum
for the high surface brightness regions (LW3 $\ge$ 20 MJy sr$^{-1}$)
in the ring is also included.
In Figure 6c, the spectrum of the lower surface brightness
disk
is given 
(6 MJy sr$^{-1}$ $\le$ LW3 $\le$ 12 MJy sr$^{-1}$).
The ring 
and extended emission each
produce $\sim$1/4 of the total mid-infrared flux
from NGC 7331, while the bulge contributes only $\sim$2$\%$.
The spectrum of the ring is similar to the global
spectrum. Furthermore, this spectrum does not change
dramatically with a higher or lower surface brightness cutoff 
(Figures 6a and 6c).
This lack of spatial variation in the mid-infrared
spectrum has important implications for the nature
of the dust emission and its relationship to
the optical/UV interstellar radiation field.
The UIB features are not significantly weaker in 
the ring than elsewhere,
thus there is no
large-scale destruction of the UIB carriers in
this region.
The comparison with the Antennae galaxy
indicates that
radiation from the molecule-sized UIB-emitting grains 
and the larger grains that produce
the 15 $\mu$m flux 
are not related to the local star formation rate in
the same way; perhaps a stronger and/or harder
optical/UV radiation field is needed to excite the grains
that produce the
15 $\mu$m radiation than the UIB carriers. 
The
UV field in the NGC 7331 ring is
not energetic enough to excite much emission
from the larger grains that dominate the 15 $\mu$m radiation,
unlike
in Knot A in the Antennae (Figure 4b).
The 15 $\mu$m surface brightness in the ring never reaches
the levels seen in Knot A ($\le$30 MJy sr$^{-1}$
in the NGC 7331 ring
vs. 230 MJy sr$^{-1}$ in Knot A (\cite{t97}).

The only place in NGC 7331 where a significant difference
in the mid-infrared spectrum is observed is
in the bulge, where
an excess in the LW2
(6.75 $\mu$m) filter is seen, relative to the flux
at the other wavelengths 
(Figure 6b).
This short wavelength excess may be due to 
photospheric light
from evolved stars.  
Assuming that 
the 2.2 $\mu$m light comes from 
3000-4000K stars,
and scaling 
the 23$''$ aperture K photometry 
of \markcite{a77}Aaronson (1977) 
to the bulge region
using the 
\markcite{p89}Pogge (1989) red continuum image, 
interpolation to the mid-infrared shows that most of the emission at 6.75 $\mu$m
can be accounted for by photospheric emission (Figure 6c).
At longer wavelengths, this proportion drops off.
At 12 $\mu$m, roughly half of the observed bulge light may
be photospheric.
A similar conclusion was reached for the bulge of 
M31, based on the IRAS data (\cite{s86}).
For the ring and diffuse emission, photospheric
contributions are negligible.

The detection of the 11.3 $\mu$m feature in the bulge,
however,
indicates that
interstellar or circumstellar dust emission is also important in
the mid-infrared.
Distinguishing between these two possibilities is difficult
because the
mid-infrared spectra
are similar (e.g., \cite{b96}; \cite{m96}).  
Scaling from the bulge of the Milky Way, \markcite{s86}Soifer
et al. (1986) conclude that in the M31 bulge there are sufficient stars
with dust shells to account for
the observed IRAS 12 $\mu$m emission.
This argument may also apply to the bulge of NGC 7331; it
has a similar mass and M/L$_{optical}$ ratio
as the M31 bulge (\cite{k87}), however, its 12 $\mu$m luminosity
is $\sim$3 times higher and its F$_{12 {\mu}m}$/F$_{2.2 {\mu}m}$
ratio is $\sim$2 times higher than in the M31 bulge (\cite{s86}),
suggesting higher contributions from interstellar dust.
On the other hand,
NGC 7331 may have a somewhat younger bulge population
than M31 (\cite{ts97}), affecting the abundance of
stars with dust shells, so this result is uncertain.  
Further
population studies are
needed to determine whether there are sufficient
dust shell stars in the NGC 7331 bulge to account for
the mid-infrared emission.

It is also unclear at present whether
sufficient interstellar gas exists in the bulge to
account for the observed 11.3 $\mu$m emission.
The IRAS data imply a 
total dust mass for NGC 7331 of 2.5 $\times$ 10$^7$ M$_{\sun}$
(\cite{y89}). 
This is likely a lower limit, since IRAS was not sensitive to very cold
dust.  
Assuming a constant mid- to far-infrared spectrum
across the galaxy, the ISO data indicate
M$_{dust}$
$\ge$ 10$^5$ M$_{\sun}$ 
in the bulge, assuming that $\sim$1/2 of the observed
bulge flux is interstellar.
Including flux missed by the interferometer and assuming
a Galactic
I$_{CO}$/n$_{H_2}$ ratio,
the CO map of \markcite{ts97}Tosaki $\&$ 
Shioya (1997) gives an upper limit of
M$_{H_2}$ $\le$ 9.7 $\times$ 10$^7$ M$_{\sun}$ 
for the bulge.  This is sufficient to account 
for the dust if the gas/dust ratio
is $\le$970, consistent with typical gas/dust ratios determined from
IRAS fluxes (\cite{y89}).
Thus it is possible that sufficient molecular gas is present in
the NGC 7331 bulge, but is fainter than current detection limits.
In contrast, the HI and H~II mass limits appear to
be too small to account for this dust.  
The HI column density 
in the center of NGC 7331 is $\sim$4 M$_{\sun}$~pc$^{-2}$ 
(\cite{b81}), implying only 
4 $\times$ 10$^6$ M$_{\sun}$ of HI in the bulge.
From the  
\markcite{p89}Pogge
(1989) image, 
L(H$\alpha$+[N~II]) $\sim$ 1.3 $\times$ 10$^6$ L$_{\sun}$
for the bulge.
This flux is mainly
[N~II], since H$\alpha$ is in absorption 
(\cite{k83c}; \cite{ot96}).
Assuming
an absorption-corrected H$\alpha$/[N~II]
ratio of 2.5 (\cite{ot96}), and assuming 10$^4$ K gas with 
n$_H$ $\ge$ 100 cm$^{-3}$ 
and A$_{\rm V}$ $\sim$ 1 (derived from the dust mass as in 
\markcite{sh96}Smith $\&$ Harvey 1996), M$_{H~II}$ 
$\le$ 6.0 $\times$ 10$^4$ 
M$_{\sun}$,
much too 
low to account for the dust.  
Thus, if they
are indeed interstellar, the very small grains
in the bulge, and the
expected larger dust grains that 
dominate the emission at the longer IRAS bands,
are likely associated with molecular gas rather than atomic
or ionized gas.
More sensitive CO measurements
are needed to test this hypothesis.

As noted above, $\sim$1/3 of the total IRAS 12 $\mu$m
flux is arising from the UIB features.
Although significant, this flux alone is not sufficient to account
for the 
F(12)/F(60) vs. L$_{60}$ trend seen in IRAS broadband colors
(\cite{s87}; \cite{hrs91}).  For moderate luminosity galaxies like NGC 7331,
F(12)/F(60) $\sim$ 0.1; this ratio decreases by a factor of 3 in high
luminosity galaxies (\cite{s87}).   
This implies that, if small grain depletion is responsible for
the observed trend in the IRAS colors, the mid-infrared continuum
as well as the emission features is affected.
Alternatively, other processes may conspire to change the 
infrared spectrum of galaxies, such as increased 10 $\mu$m absorption
in high luminosity galaxies or variations in large grain temperatures
or properties.

\section{Conclusions}

ISOCAM imaging of the Sb galaxy NGC 7331 has revealed
a bright ring encircling
an extended central bulge.  
The mid-infrared spectrum of NGC 7331
shows one of the first detections of the 
7.7 and 11.3 $\mu$m dust emission features in
a quiescent spiral galaxy.
The continuum at 15 $\mu$m is weak
with respect to the UIB features, in contrast to starburst galaxies, consistent
with the moderate L$_{FIR}$ of NGC 7331 and its classification
as a relatively quiescent spiral galaxy.
Little variation in the mid-infrared spectrum across the galaxy
is found with the exception of the
bulge, which
shows excess 6.75 $\mu$m emission, probably caused by
photospheric emission.
The 11.3 $\mu$m feature is also detected in the bulge,
however,
indicating additional contributions
from interstellar or circumstellar dust.

\acknowledgements

I am grateful to the ISO team
for making this project possible.
I especially thank the ISOCAM intrument team
and the ISO staff at IPAC for their help.
This research has made use of the NASA/IPAC Extragalactic
Database (NED) which is operated by the Jet Propulsion Laboratory
under contract with NASA.
This work was supported by ISO data analysis funding from NASA.

\vfill
\eject

{\bf Captions}

\figcaption{(Plate L1) The Digitized Sky Survey optical image of NGC 7331.
The region which was mapped by ISOCAM is outlined.
The three small galaxies to the east of NGC 7331 are, south to north,
NGC 7337, NGC 7335, and NGC 7336.  NGC 7337 and NGC 7335 have redshifts
well in excess of that of NGC 7331 (de Vaucouleurs et al. 1991;
Wegner, Haynes, $\&$ Giovanelli 1993), indicating that they
are background galaxies.  NGC 7336 has no published velocity.
The Digitized Sky Survey, a compressed
digital form of the Palomar Observatory Sky Atlas,
was produced at the Space Telescope Science Institute
under U.S. Government grant NAG W-2166.}

\figcaption{(Plate L2) The six mosaicked ISO maps of the central region of NGC 7331.
For Figures 2a-f, the filters are
LW2, LW3, LW6, LW7, LW8, and LW9, respectively. The region mapped by ISO
is 5\farcm8 $\times$ 2\farcm25.  The orientation is marked.  Note that the  
bad ISOCAM column (Cesarsky et al. 1996) 
is visible at the edges of the mosaicked image, where no overlap is available.
For the LW6-9 filters, additional data points are missing because of vignetting
at the edges of the array.}

\figcaption{(Plate L3) The inner region of NGC 7331 at five wavelengths.
a) The narrowband 6435$\AA$ continuum image from Pogge (1989).
b)
The H$\alpha$+[N~II]
map from Pogge (1989).
c) The LW3 map.
d) The
Pogge (1989) H$\alpha$+[N~II] map (contours), 
superposed on the LW3 map.
e) The 20 cm radio continuum map from Cowan et al. (1994) (contours),
superposed on the LW3 map.
f) The Nobeyama CO (1 $-$ 0) map
(Tosaki $\&$ Shioya 1997) (contours),
superposed on the LW3 map.
The field of view of these images
is 88$''$ $\times$ 152$''$.  North is up and east is to the left.
The spatial resolution in the radio
continuum and CO maps is 1\farcs8 $\times$ 1\farcs4 and 7\farcs3
$\times$ 3\farcs7,
respectively.  
The primary beam of the Nobeyama array is 65$''$; beyond this region
the CO map is not reliable.
The ISO maps have been aligned using the location of the
nucleus on the H$\alpha$+[N~II] maps.}

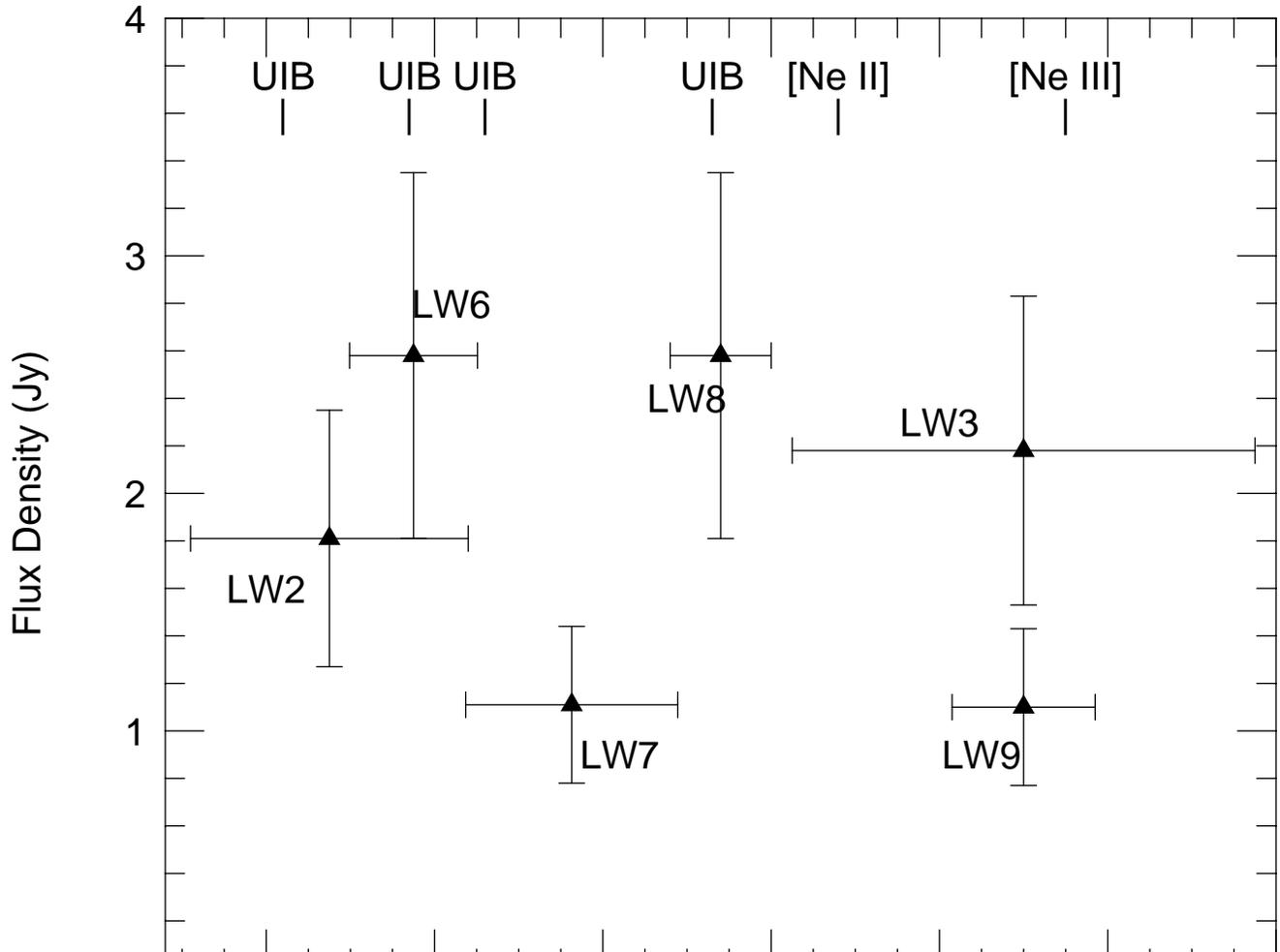
\figcaption{a). The mid-infrared spectral energy distribution
of the mapped region of NGC 7331 (filled triangles).
Flux uncertainties of 30$\%$ are shown.
The horizontal error bars give the wavelength extent of the filters.
The six filters are identified.
b) The mid-infrared spectrum for the sky near NGC 7331
(filled triangles).  
The zodiacal spectrum, as measured by the ISOCAM CVF (Reach
et al. 1996, revised by Reach 1997), is also plotted (dotted line).
The left axis gives the surface brightness scale for the NGC 7331
sky values; the right axis is the scale for the Reach (1997) data.
Blackbody curves for 262K and 280K are also displayed (solid and dashed,
respectively).}

\figcaption{The NGC 7331 spectrum, compared with
the ISOCAM CVF data for two regions in the
Antennae galaxies NGC 4038/4039 (Vigroux et al. 1996, as updated
by Tran 1997).
The solid line is the spectrum for Knot A, a luminous starburst
region, while the dotted line is for the nucleus of NGC 4038.
The prominent dust features and the 12.8 $\mu$m [Ne~II] and 15.5 $\mu$m [Ne~III]
features are marked.}  

\figcaption{The mid-infrared spectrum for a) the NGC 7331 ring, b) the
NGC 7331 bulge, and c) the lower surface brightness emission
in NGC 7331.  In Figures 6a and 6b, data for regions with LW3 surface
brightness greater than 12 MJY sr$^{-1}$ are plotted as filled
triangles.  In Figure 6a, data for regions with LW3 surface brightness greater
than 20 MJy sr$^{-1}$ are given as open circles.  In
Figure 6b, the solid and dashed lines are interpolations
from the near-infrared K band flux density for the bulge, assuming blackbody
radiation at temperatures of 3000K and 4000K, respectively.}


\begin{thebibliography}{}

\bibitem[Aaronson 1977]{a77} Aaronson, M. 1977, Ph.D. Thesis, Harvard University

\bibitem[Borghesi, Bussoletti, $\&$ Colangeli 1987]{b87}
Borghesi, A., Bussoletti, E., $\&$ Colangeli, L. 1987, \apj, 314, 422 

\bibitem[Bosma 1981]{b81} Bosma, A. 1981, \aj, 86, 1791

\bibitem[Boulanger $\&$ Perault 1988]{bp88} Boulanger, F., $\&$ 
Perault, M. 1988, \apj, 330, 964

\bibitem[Boulanger et al. 1996]{b96} Boulanger, F., et al. 1996, \aap, 315, L325

\bibitem[Bronfman et al. 1988]{bron88}Bronfman, L.,
Cohen, R. S., Alvarez, H., May, J., $\&$ Thaddeus,
P. 1988, \apj, 324, 248

\bibitem[Cesarsky et al. 1996]{c96} Cesarsky, C. et al. 1996, \aap, 315, L32

\bibitem[Cowan et al. 1994]{c94}Cowan, J. J., 
Romanishin, W., $\&$ Branch, D. 1994,
\apj, 436, L139

\bibitem[Cox, Kr\"ugel, $\&$ Mezger 1986]{c86}Cox, P., Kr\"ugel, E., $\&$
Mezger, P. G. 1986, \aap, 155, 380

\bibitem[de Vaucouleurs et al. 1991]{RC3}de 
Vaucouleurs, G., de Vaucouleurs, A., Corwin, Jr., H. G.,
Buta, R. J., Paturel, G., $\&$ Fouque, R. 1991, Third Reference
Catalogue of Bright Galaxies, Version 3.9

\bibitem[Gautier et al. 1984]{g84}Gautier, T. N., et al.
1984, \apj, 278, L57

\bibitem[Habing et al. 1984]{h84}Habing, H. J., et al. 1984, \apj,
278, L59

\bibitem[Helou et al. 1991]{hrs91} Helou, G., Ryter, C., $\&$ Soifer, B. T. 1991, \apj, 376, 505

\bibitem[Helou et al. 1996]{h96} Helou, G., et al. 1996, \aap, 315, L157

\bibitem[Keel 1983a]{k83a} Keel, W. C. 1983a, \apjs, 52, 229

\bibitem[Keel 1983b]{k83b} Keel, W. C. 1983b, \apj, 268, 632

\bibitem[Keel 1983c]{k83c} Keel, W. C. 1983c, \apj, 269, 466

\bibitem[Kent 1987]{k87} Kent, S. M. 1987, \aj, 93, 816

\bibitem[Kessler et al. 1996]{k96} Kessler, M. F., et al. 1996, \aap, 315, L27

\bibitem[L\'eger $\&$ Puget 1984]{lp84} L\'eger, A., $\&$ Puget, J. L. 1984, \aap, 137, L5

\bibitem[Molster et al. 1996]{m96}Molster, F. J. et al. 1996, \aap, 315, L373

\bibitem[Ohyama $\&$ Taniguchi 1996]{ot96}Ohyama, Y., $\&$ Taniguchi, Y.
1996, in The Physics of LINERS in View of Recent Observations,
ASP Conference Series, Vol 103, p. 205

\bibitem[Papoular et al. 1989]{pap89}Papoular, R. et al. 1989, \aap, 217, 204

\bibitem[Pogge 1989]{p89}Pogge, R. W. 1989, \apjs, 71, 433

\bibitem[Reach et al. 1996]{r96}Reach, W. T. et al. 1996, \aap, 315, L381

\bibitem[Reach 1997]{r97}Reach, W. T. 1997, private communication

\bibitem[Rice et al. 1988]{r88}Rice, W., Lonsdale, C. J., Soifer, B. T.,
Neugebauer, G., Kopan, E. L., Lloyd, L. A., de Jong, T., $\&$ Habing, H. J.
1988, \apjs, 68, 91

\bibitem[Roche et al. 1991]{r91} Roche, P., Aitken, D., Smith, C., $\&$ Ward, M. 1991, \mnras, 248, 606

\bibitem[Sakata et al. 1984]{sak84}Sakata, A. et al. 1984, \apj, 287, L51

\bibitem[Sanders, Solomon, $\&$ Scoville 1984]{sanders84}Sanders, D. B.,
Solomon, P. M., $\&$ Scoville, N. Z. 1984, \apj, 276, 182

\bibitem[Sauvage et al. 1996] {sau96}Sauvage, M. et al. 1996, \aap, 315, L89

\bibitem[Sellgren 1984] {s84} Sellgren, K. 1984, \apj, 277, 623

\bibitem[Seibenmorgen et al. 1997]{s97}Siebenmorgen, R., Starck, J. L.,
Cesarsky, D. A., Guest, S., $\&$ Sauvage, M. 1997, ISOCAM Data User's Manual,
Version 3.0

\bibitem[Smith et al. 1987] {s87} Smith,
B. J., Kleinmann, S. G., Huchra, J. P.,
$\&$ Low, F. 1987, \apj, 318, 161

\bibitem[Smith $\&$ Harvey 1996] {sh96} Smith, B. J., $\&$ Harvey, P. M. 1996, \apj, 468, 139

\bibitem[Sodroski et al. 1997]{sodroski97}Sodroski, T. J., Odegard, N., Arendt, R. G.,
Dwek, E., Weiland, J. L., Hauser, M. G., $\&$ Kelsall, T. 1997,
\apj, 480, 173

\bibitem[Soifer et al. 1986]{s86} Soifer, B. T., Rice, W. L, Mould, J. R.,
Gillett, F. C., Rowan-Robinson, M., $\&$ Habing, H. J. 1986, \apj, 304, 651

\bibitem[Starck et al. 1995]{s95} Starck, J. L., Bijaoui, A., $\&$ Murtagh, F.
1995, `Multiresolution Support Applied to Image Filtering
and Deconvolution'', in CVIP: Graphical
Models and Image Processing, Vol. 57, 420

\bibitem[Tosaki $\&$ Shioya 1997]{ts97} Tosaki, T., $\&$ Shioya, Y. 1997, \apj, 
484, 664

\bibitem[Tran 1997]{t97}Tran, D. 1997, private communication

\bibitem[Vigroux et al. 1996]{v96} Vigroux, L., et al. 1996, \aap, 315, L93

\bibitem[von Linden et al. 1996]{von96} von Linden, S., Reuter, H.-P., Heidt, J., Wielebinski, R.,
$\&$ Pohl, M. 1996, \aap, 315, 52

\bibitem[Wegner et al. 1993]{whg93} Wegner, G., Haynes, M. P., $\&$ Giovanelli,
R. 1993, \aj, 105, 1251

\bibitem[Young et al. 1989]{y89} Young, J. S., Xie, S., Kenney,
J. D. P., $\&$ Rice, W. L. 1989, ApJS, 70, 699


\end{thebibliography}
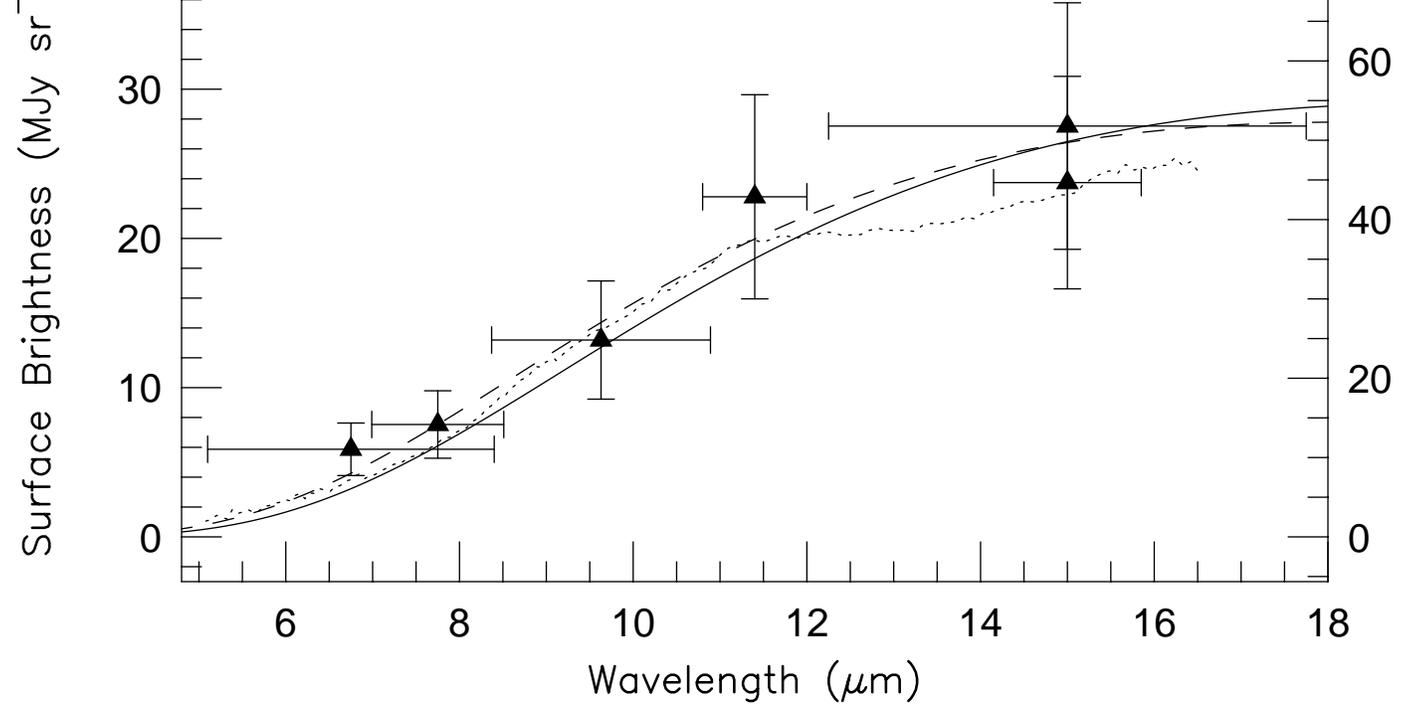
\end{document}